# A Kind of Boundary-Layer 'Flutter': The Turbulent History of a Fluid Mechanical Instability

Michael Eckert

Abstract:

The transition from laminar to turbulent flow has been a notorious riddle in fluid dynamics since the nineteenth century. Hydrodynamic instabilities were regarded as a cause for the onset of turbulence, but their theoretical investigation led to paradoxical results. I focus on studies about the instability of laminar boundary-layer flow along a flat plate. The experimental discovery of boundary-layer oscillations ("Tollmien-Schlichting waves") in the USA during World War II vindicated a long-disputed theory developed in Germany in the 1920s and '30s. However, the instability which causes these oscillations marks only the initial phase of the onset of turbulence; higher-order instabilities had to be assumed in order to extend the analysis to the regions of the transition zone where more complex phenomena (such as "turbulent spots") mark the onset of turbulence. There is still no theory to describe these processes in a satisfactory manner. The emphasis on the fundamental character of the problem, however, tends to obscure the practical interests which motivated the study of boundary-layer instabilities throughout its long history. The disputed theories played a role in the quest for "laminar airfoils" and other practical applications. Theory and practice are entangled more intimately than it would appear from the rhetoric about the fundamental character of the "turbulence problem".

Hydrodynamic stability has been a major concern in fluid mechanics since nineteenth-century physicists like Hermann Helmholtz, William Thomson (Lord Kelvin), John William Strutt (Lord Rayleigh), and others exposed some of the difficulties involved for a theoretical description (Darrigol 2005: chap. 5). Instabilities were regarded as a cause for the onset of turbulence.[1] However, each flow configuration (pipe flow, flow along plane and curved walls, stratified flow, flow between rotating cylinders etc.) demanded separate treatment and displayed different kinds of instabilities. Despite considerable efforts throughout the twentieth century, no general solution was accomplished. Even for simple configurations, such as plane parallel flows, models that are successful in one situation "may fail dismally in another", as a reviewer remarked in the year 2000 (Bowles 2000: 246). Decades of sophistication in applied mathematics and theoretical physics only served to reformulate the problem now and then in new guises. What appears simple in principle—to predict the onset of turbulence from the breakdown of laminar flow—crystallized as a major challenge for theoretical fluid dynamics. More than a century of theoretical inquiry into this phenomenon could not alter the observation that "the instability of parallel flows of a viscous fluid is notoriously subtle" (Drazin 2002: 160).

---

[1] The riddle of turbulence as perceived at the beginning of the 20th century involves the onset of turbulence and fully developed turbulence. This article deals only with the former; for a broader discussion including the latter see: Eckert 2008a.

In retrospect, the frustrated efforts to cope with turbulence tend to be mystified if associated only with the names Kelvin, Rayleigh, Sommerfeld, Heisenberg, Feynman, and other outstanding theorists. Theodore von Kármán, for example, recalled in his autobiography that Arnold Sommerfeld, "the noted German theoretical physicist of the 1920s", once told him "that before he died he would like to understand two phenomena—quantum mechanics and turbulence." In von Kármán's opinion from the 1960s, Sommerfeld was "somewhat nearer to an understanding of the quantum, the discovery that led to modern physics, but no closer to the meaning of turbulence" (Kármán 1967: 134; see also Gorn 1992: 61). This paper attempts to contextualize the experimental and theoretical investigations about boundary-layer instability in order to unearth the more practical concerns involved with these studies. I will start with a brief survey of instability theories in the wake of Rayleigh's pioneering efforts around 1880. With regard to boundary-layer flow, the climax was reached in the early 1930s with the so-called Tollmien-Schlichting theory—a theory which described the effect of the instability as an amplification of infinitesimal disturbances such that plane waves of certain wavelengths would emerge. However, according to the majority view, the onset of turbulence in boundary layers was induced by finite external disturbances rather than as a consequence of an intrinsic instability. Wind-tunnel tests provided support for this view: The transition from laminar to turbulent flow of test models in wind tunnels was found to depend on the degree of turbulence of the air stream. By reducing wind-tunnel turbulence the length of laminar flow along the chord of an airfoil could be increased. If airfoils could be shaped such that the boundary layer were laminar for the major part along the chord—such airfoils were called "laminar airfoils"—then the skin friction was reduced. Thus, by the late 1930s, low-turbulence wind tunnels and laminar airfoils became subjects of research in aeronautical laboratories in many countries. Ironically, the Tollmien-Schlichting theory, which was disputed in view of wind-tunnel experiments, was vindicated in this same context. When aerodynamicists at the National Bureau of Standards investigated the flow along a flat plate in a low-turbulence wind tunnel by means of hot-wire anemometry during World War II, they discovered—against their own expectations—boundary-layer oscillations. After the war, hydrodynamic stability theory aroused new interest and at the same time became the subject of new controversies. In the 1950s, new experimental discoveries, such as "turbulent spots", showed that the onset of turbulence was more complicated; Tollmien-Schlichting waves merely characterized the initial phase of the transition process. By conceiving nonlinear theories (the Tollmien-Schlichting theory described the amplification of infinitesimal disturbances in a linear approximation) it was hoped to extend the reach of stability theory and to take finite disturbances into account. Nonlinear theories were flourishing in the 1960s. But the "expansive sixties", as this phase of research was recalled from a theoretical perspective, rather exposed new difficulties instead of settling the problem.

This is, in a nutshell, the skeleton of the following narrative. However, we have to add the details from the historical records—letters, reports, original publications—in order to see how theory and practice became entangled in this history. The "great mystery"-rhetoric of the turbulence riddle tends to obscure the character of this research: Its portrayal as a quest to solve the last riddles of classical physics is a retrospective embellishment. The historical contextualization reveals a closer relation with down-to-earth problems than such embellishments suggest. But no simple shift in the study of hydrodynamic stability and turbulence closer to the "applied" end of the range of "pure" and "applied" research topics will suffice for the purpose of historical analysis. The entangled theory–practice relation in this case, as in others, cannot be described by a "linear model"[2] of science–technology relations. The turbulent historical dynamics of this research field, therefore, is also a challenge for the historiography of science and technology.

---

[2] According to the so-called "linear model", technology is assumed to result from science in a one-way process, from the "basic" towards the "applied". For recent discussions of the long-debated issue of science—technology relationship see: Grandin et al. 2005, Trischler and Eckert 2005, Forman 2007.

**The Origins of the Tollmien-Schlichting Theory**

The "turbulence problem" as perceived early in the twentieth century should not be regarded as one and the same as the more general perception of turbulence as a fundamental problem of classical physics. The latter includes the problem of fully developed turbulence, i. e., chaotic turbulent flow without regard for its onset; the former addressed the transition to turbulence, i. e., the breakdown of ordered flow before it becomes fully chaotic. Developed turbulence was made the subject of statistical analysis and (in recent decades) chaos theory. The onset of turbulence was regarded as a case of flow instability. However, this is merely a rough conceptual distinction. Neither historically nor from the contemporary vantage point of physics, mathematics, or engineering, may turbulence analysis be sorted into either one or the other category exclusively. What has been addressed at one time or another as "the turbulence problem" should rather be perceived as a bundle of problems that adapted to changing epistemic environments (Eckert 2008a).

This study is concerned with the turbulence problem as it was explicitly articulated at the beginning of the twentieth century, i. e., as a riddle that belongs to the category of flow instability. The principle of hydrodynamic stability analysis, as conceived by William Thomson (Lord Kelvin), John William Strutt (Lord Rayleigh) and others, was as simple as the mathematics to pursue it was difficult: Superpose a wavelike infinitesimal disturbance upon a two-dimensional parallel flow with a given velocity profile and see whether it grows or decays. If the disturbance grows, the flow is unstable; if it decays, the flow is stable. Rayleigh derived a theorem according to which instability requires an inflection of the velocity profile (Darrigol 2005: 209–10). The approach was later extended by William McFadden Orr and

Arnold Sommerfeld (independently from each other) for the viscous case. They derived an equation for a plane flow with a linear velocity profile ("Couette flow") which offered the prospect of determining a critical Reynolds number at which the flow would become unstable (Orr 1907, Sommerfeld 1908). Richard von Mises, however, published a proof in 1912 that if plane Couette flow was stable within the limit of vanishing Reynolds numbers—which was demonstrated by Sommerfeld—then it must be stable for all Reynolds numbers (Mises 1912). By then Sommerfeld's student, Ludwig Hopf, solved the "Orr-Sommerfeld equation"—and found no limit of stability (Hopf 1914). Plane Couette flow, therefore, would never cease to be laminar according to the Orr-Sommerfeld approach—in obvious contradiction to practical experience (Eckert 2010).

From 1916, Ludwig Prandtl pursued a "working program on turbulence theory" that also involved the study of flow instabilities as a cause for the onset of turbulence (Fig. 1; for more detail, see Eckert 2006: chap. 5.2).

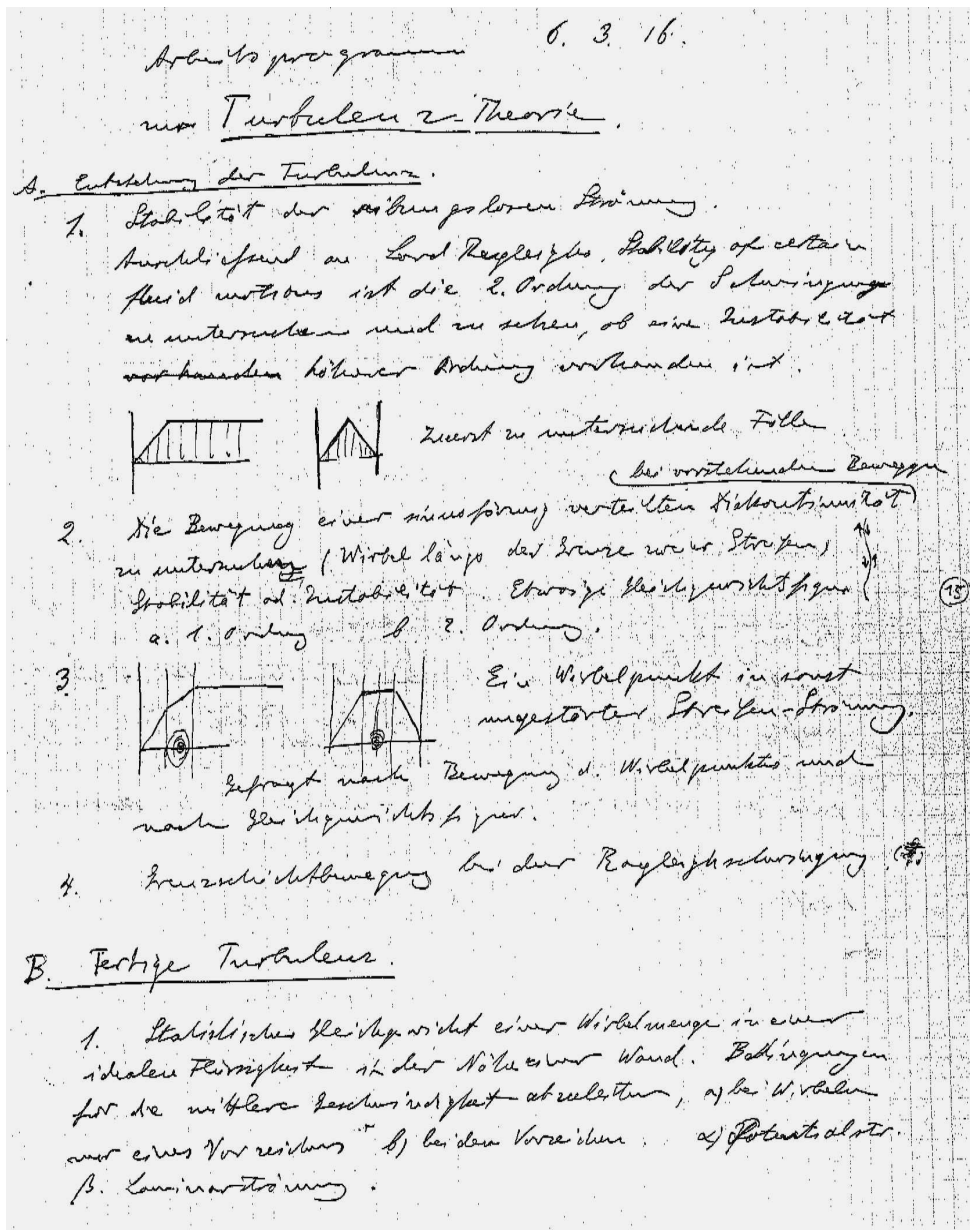

Fig. 1: In this draft[3] Prandtl discerned "A. Onset of turbulence" and "B. Developed turbulence" as the two major problem areas for his research program on the theory of turbulence. "A" involved the analysis of hydrodynamic stability of laminar flows with given velocity profiles; Rayleigh's investigation of the stability of inviscid flows with piecewise linear velocity profiles served as a starting point, from which Prandtl attempted to analyse the stability of laminar boundary layer flow (Blasius flow).

He attempted to apply the stability analysis to the flow in a boundary layer. If the velocity profile of plane laminar boundary-layer flow (Blasius profile) was approximated by piecewise straight lines, Rayleigh's inviscid theory would predict stability because there would be no inflexion. Given the previous experience of the Orr-Sommerfeld analysis, Prandtl assumed that adding viscosity would increase the stability. However, when Prandtl's doctoral student, Oskar Tietjens, elaborated the theory, the profile was found to be unstable even for very low Reynolds numbers, "contrary to the dogma" (Prandtl 1922: 692). But this analysis did not solve the problem. Where previous studies failed to determine an onset of instability, the Prandtl-Tietjens approach failed to account for stability. Both theoretical results were in blatant conflict with experimental observations, which showed that boundary-layer flow was stable for low Reynolds numbers and unstable for high ones. Tietjens believed that the approximation of a piecewise linear velocity profile for the Blasius flow was responsible for the "unsatisfactory result" of his doctoral thesis (Tietjens 1925: 214).

---

[3] Manuscript „Turbulenz I" Göttingen, Niedersächsische Staats- und Universitatsbibliothek (NSUB), Teilnachlass Ludwig Prandtl, Acc. Mss. 1999.2, Cod. Ms. L. Prandtl 18, Blatt 15.

At about that time Werner Heisenberg made another attempt to predict the onset of turbulence in his doctoral dissertation under Sommerfeld. Heisenberg analysed plane Poiseuille flow (where the velocity profile is parabolic) and found a limit of stability (Heisenberg 1924). This was another case "against the dogma" of stability for profiles without inflection. However, Fritz Noether, another Sommerfeld pupil, denied in his fundamental critique of previous stability theories that Heisenberg's result was valid (Noether 1926). Noether's analysis provoked a controversy with Prandtl, too, but no agreement was reached (Eckert 2008a: 44–6). Flow instability became a bone of contention for mathematically oriented fluid dynamicists and theoretical physicists. "The history of the solutions of the Sommerfeld-Orr equations," Heisenberg recalled later, was "rich in ups and downs of surprises, errors, drawbacks". Even with the hindsight of decades of research, Heisenberg admitted, "it is not known what is actually wrong in the work of Noether" (Heisenberg 1969: 47).

The dispute with Noether motivated Prandtl to launch a new effort. He suggested the analysis of boundary-layer stability as the subject matter for another doctoral dissertation. The weak point of Tietjens's theory, the piecewise linear velocity profile, should be replaced by a more realistic profile closer to the Blasius profile for the laminar flow in the boundary layer along a flat plate. After the first doctoral student failed,[4] Walter Tollmien was entrusted with this problem. Tollmien found as a result a rather complex distribution of stable and unstable states of flow, depending both on the Reynolds number and the wavelength of the disturbance superposed on the laminar flow. He obtained a critical Reynolds number of 420 as the lower limit beyond which a disturbance could be amplified. Although this was much lower than the Reynolds numbers at which the onset of turbulence was experimentally observed, the discrepancy appeared plausible because at this lowest Reynolds number only one disturbing wave at a certain wavelength would be allowed to grow; the likelihood of a breakdown in the stability would increase only at higher Reynolds numbers where more wavelengths were affected (Tollmien 1929).

During the subsequent years, Prandtl made stability analysis the theme of more research work at his institute—theoretical research as well as experimental. Hermann Schlichting applied Tollmien's procedure to the flow in a rotating cylinder and revised the stability analysis of plane Couette flow. For the latter case he showed that only if the linear velocity profile was assumed from the very beginning, as in Hopf's and von Mises's theories, would this flow be stable for all Reynolds numbers. If, however, the starting process was taken into account, during which the velocity profile gradually assumed linear form, this flow displayed a limit of stability, too. In 1933, Schlichting further elaborated the stability analysis of plane boundary-layer flow in order to facilitate comparison with experiments. In Tollmien's doctoral dissertation only the limits of stability had been determined; Schlichting's extension resulted in estimates of the size to which unstable disturbances would grow (Schlichting 1933 and 1935). Experimental evidence for the unstable modes was sought using a 6-m-long and 20-cm-wide water channel with holes in the wall, by which the flow in the boundary layer could be modified, but the evidence was inconclusive (Nikuradse 1933, Prandtl 1933, Schlichting 1934).

---
[4] Prandtl to Hopf, 20 July 1926. MPGA, div. III, rep. 61, no. 704.

**Opposing Views Based on Wind-tunnel Turbulence**

"The stability of the Blasius flow has been analysed by Tollmien by an approximate method which is open to some analytical criticism", Geoffrey Ingram Taylor remarked in 1936. He offered experimental data which seemingly contradicted the stability approach: According to these data the boundary layer became unstable only at much higher Reynolds numbers than predicted by Tollmien. "This fact constitutes a serious criticism of Tollmien's criterion", Taylor concluded. As long as there was no tangible evidence for the Tollmien-Schlichting theory, an alternative view of the laminar–turbulent transition seemed more in accordance with available wind-tunnel data: Even without the disputed amplification of infinitesimally small disturbances, the laminar boundary layer would be forced into a transition to turbulence as a consequence of other disturbances than those of the Tollmien-Schlichting type. All available evidence, so far, suggested that "the change point from steady to turbulent flow depends only on the turbulence of the stream ... It seems that the way in which turbulence is most likely to affect the boundary layer is through the action of local pressure gradients which necessarily accompany turbulent flow." (Taylor 1936: 307–8).

    The same conclusion was drawn at the Aerodynamics Section of the National Bureau of Standards, where Hugh Dryden directed a series of wind-tunnel tests on the influence of turbulence. By "numerous experiments on the critical Reynolds Number of spheres and airship models in relation to measurements of the fluctuations present in the air stream", Dryden became convinced "that it is the amplitude of the disturbances initially present, rather than their frequency" which causes the transition to turbulent flow in the boundary layer (Dryden 1936: 344).

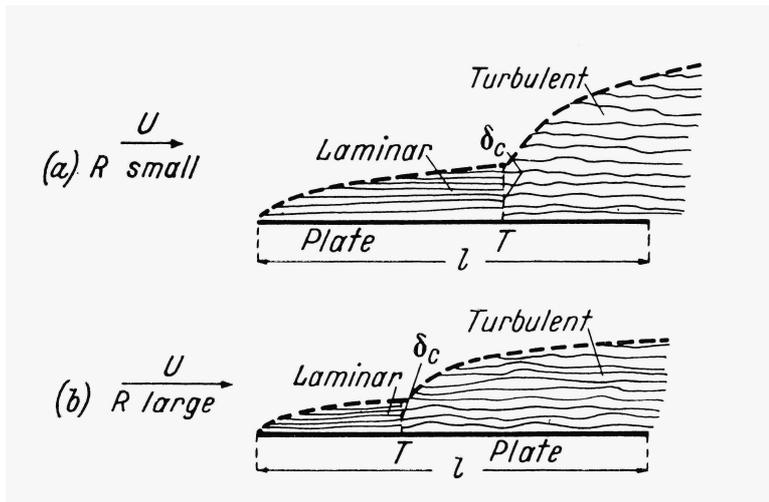

Fig. 2: The transition from laminar to turbulent flow in the boundary layer depends on the Reynolds number of the airstream in a wind tunnel. Skin friction is lower for laminar boundary layer flow; therefore, the location of the transition T affects the drag and lift measurements in wind tunnels. For this reason, investigations about the onset of turbulence became important research topics of aeronautical laboratories (Kármán & Millikan 1934: Figure 6).

Wind-tunnel turbulence had become a serious concern for aeronautical research because it affected the trust in model testing (Fig. 2). Even if the same models were tested in various tunnels at the same Reynolds number, the results differed due to different degrees of turbulence. Model tests at the NACA's Langley Aeronautical Laboratory performed, for example, in the small Variable Density Tunnel (VDT) and the large Full-Scale Tunnel (FST), differed considerably. Several efforts were made to find corrections that accounted for tunnel turbulence. Consolation was sought in the observation that the boundary layer in free flight at high Reynolds numbers would be turbulent anyway. But it turned out that free-flight turbulence was different from wind-tunnel turbulence. The disputes about the onset of turbulence in boundary layers, therefore, touched both theoretical and practical concerns: on the one hand, they had a bearing on the epistemic perception of the turbulence problem (Eckert 2008a: 63–5), on the other, they were related to very practical aeronautical concerns —concerns that became manifest in further sophistication in wind tunnels.[5]

---

[5] The dual role of wind tunnels as engineering devices and research instruments is discussed more generally elsewhere (Eckert 2008b).

By the end of the 1930s, the impact of wind-tunnel turbulence on the lift and drag of test models was demonstrated so pervasively that other causes for boundary-layer transition seemed rather esoteric. On the occasion of the Fifth International Congress of Mechanics in Cambridge, Massachusetts, G. I. Taylor reviewed the current state of affairs about the onset of turbulence as follows: "One school has thought that it is due to a definite instability, i. e. to a condition in which infinitesimal disturbances will grow exponentially." Here he was alluding to the Göttingen approach, but went on: "Another regards the motion in most cases as definitely stable for infinitesimal disturbances but liable to be made turbulent by suitable disturbances of finite magnitude or by large enough reverse pressure gradient." Wind-tunnel testing corroborated the latter view: onset of turbulence in the boundary layer of test models was clearly related to the turbulence that was present in the airstream of a wind tunnel. Therefore, Taylor had little doubt about concluding that "the experimental evidence is against the instability theory" (Taylor 1939: 308).

**Low-turbulence Wind Tunnels and Laminar Airfoils**

In 1935, the NACA's wind-tunnel expert at the Langley Laboratory, Eastman Jacobs, returned from a trip to Europe where he had discussed with Geoffrey I. Taylor the impact of wind-tunnel turbulence on the profile drag. Taylor was at that time working on a statistical theory of turbulence and studying methods to control and quantize the degree of turbulence by means of grids and honeycombs in wind tunnels. Few were as aware as he of the impact that small-scale variations of the airstream in wind tunnels exerted on drag measurements with test models. Back in the Langley Laboratory, Jacobs urged his colleagues during an internal conference on boundary-layer control to support a proposal to construct suitable low-turbulence testing equipment. A reduction in wind-tunnel turbulence could be achieved by inserting a grid, in order to make the airstream fluctuations more homogeneous, and forcing the air through a constriction by which the ground speed was increased, so as to reduce the ratio of speed fluctuations to ground speed. Jacobs further elaborated his ideas in a paper on "Laminar and Turbulent Boundary Layer as Affecting Practical Aerodynamics". By May 1937, the NACA approved Jacobs proposal to build a new, pressurized, low-turbulence tunnel as a successor to the older Variable Density Tunnel. Officially they designated it as an "icing tunnel", because it could be used to study the causes of airplane crashes due to icing at high altitudes. But this was a subterfuge, because boundary-layer research sounded too academic for justifying the construction of such an expensive new facility. Except for some icing experiments during test runs when the facility was still under construction, the "Two-Dimensional Low-Turbulence Pressure Tunnel", as it was finally called, was never used for the announced purpose (Hansen 1987: 109–11, Doenhoff and Abbott 1947).

With the availability of low-turbulence wind tunnels, it became possible to investigate an idea which Theodore Theodorsen, the head of Langley's Physical Research Division, had conceived in the early 1930s: If the curvature of a wing's profile was shaped such that the air moved as long as possible in an accelerating pressure gradient, transition to turbulence in the wing's boundary layer could be delayed. The skin friction of such "laminar airfoils" was reduced to the extent that the point of transition to turbulence was moved away from the leading edge. By the mid-1930s, methods were developed to shape a profile according to a prescribed pressure distribution. (Traditional airfoil theory determined the pressure distribution for a given profile; the laminar airfoil problem called for a solution to the inverse problem.) The measurements in the new, low-turbulence tunnel soon showed that the drag of appropriately shaped laminar airfoils was only half of that of conventional profiles. On the eve of World War II, this result came in time for the design of faster fighter airplanes. Although the details were published only in an Advance Confidential Report, the news were not entirely kept secret. In its Annual Report for 1939, published in 1940, the NACA announced that "airfoils should be designed to take advantage of true low-drag laminar boundary layers over a greater portion of the airfoil" and revealed that "preliminary investigations were started by the development of new airfoil forms that, when tested in the new equipment [i. e., the low turbulence tunnel], immediately gave drag coefficients of one-third to one-half the values obtained for conventional sections" (NACA 1940: 10, Hansen 1987: 112–16).

Work on low-drag wings became a major part of the NACA's contribution to the war effort. Within less than two years, theoretical determination of laminar profiles and tests in the new low-turbulence tunnel resulted in the design of a low-drag wing for the P-51 Mustang, a fighter airplane produced by North American Aviation. Although the airflow in the boundary layer of Mustang wings was probably not as laminar as proclaimed, this airplane's performance was outstanding. The NACA regarded its research for the Mustang as one of its greatest contributions to the war effort. NACA's Executive Secretary, John Victory, claimed that the Germans never unravelled the superior aerodynamic features of the Mustang wing because their wind tunnel did not have the required low intensity of turbulence (Roland 1985, vol. 2: 549)

What did the Germans know about laminar airfoils? In 1940, the Lilienthal Society, an aeronautical research organisation under the umbrella of the Third Reich's Air Ministry, awarded prizes for work which dealt with the following problem: "For the further improvement of aircraft performance it is necessary to reduce the drag of all surfaces exposed to the airstream. According to the results of research on friction layers, successes may be expected if the boundary layer is kept laminar as far as possible downstream; the transition point beyond which the boundary layer is turbulent should be moved away as far as possible from the stagnation point at the leading edge." (Lilienthal-Gesellschaft 1940: 3) The prize committee, chaired by Tollmien, particularly called for investigations aimed at predicting the transition from laminar to turbulent boundary-layer flow in terms of its "dependence on the shape of the body or profile of the wing". The Lilienthal Society published the awarded research papers in a secret wartime report. The first prize was awarded to Hermann

Schlichting and his assistant, Albert Ulrich, from the Technische Hochschule (TH) in Braunschweig, for a two-part paper on calculating the laminar–turbulent transition. Based on the Tollmien-Schlichting concept of boundary-layer instability on a flat plate, they developed a numerical procedure for arbitrary wing profiles. The curvature of the surface involves pressure gradients which accelerate or slow down the flow in the boundary layer. Schlichting and Ulrich derived curves of neutral stability for various accelerations which showed how, with increasing acceleration, the stability of the laminar flow is increased—and therefore the transition point to turbulent flow moves towards the rear edge of the wing profile. In the second part they applied their procedure to profiles with the thickest part in the rear instead of in front ("Dickenrücklage"), so that the pressure gradient was most favorable for laminar flow and the transition to turbulence was delayed as far as possible. In a third part, which was submitted in 1941 after the contest but added to the prize report, they considered some consequences of such laminar profiles (Lilienthal-Gesellschaft 1940).

In October 1941, the General Fluid Dynamics Committee of the Lilienthal Society convened a conference on "Boundary-Layer Problems". On this occasion Schlichting and his collaborators presented a systematic study about the influence of the shape of a profile on the point where the laminar boundary becomes turbulent (Lilienthal-Gesellschaft 1941). Schlichting also reviewed the contemporary knowledge about the laminar–turbulent transition and the potential of laminar airfoils in a seminar at the "Luftfahrtforschungsanstalt Hermann Göring" (LFA) in Völkenrode near Braunschweig (Fig. 3). He presented a diagram which compared the drag coefficients of conventional and laminar profiles, among which some displayed exceptional low drag. However, the drag coefficients were minimal only at zero angle of attack and within a narrow range of Reynolds numbers. Beyond this range they increased sharply, which indicated a sudden move of the transition point along the chord from the rear to the leading edge (Schlichting 1941).

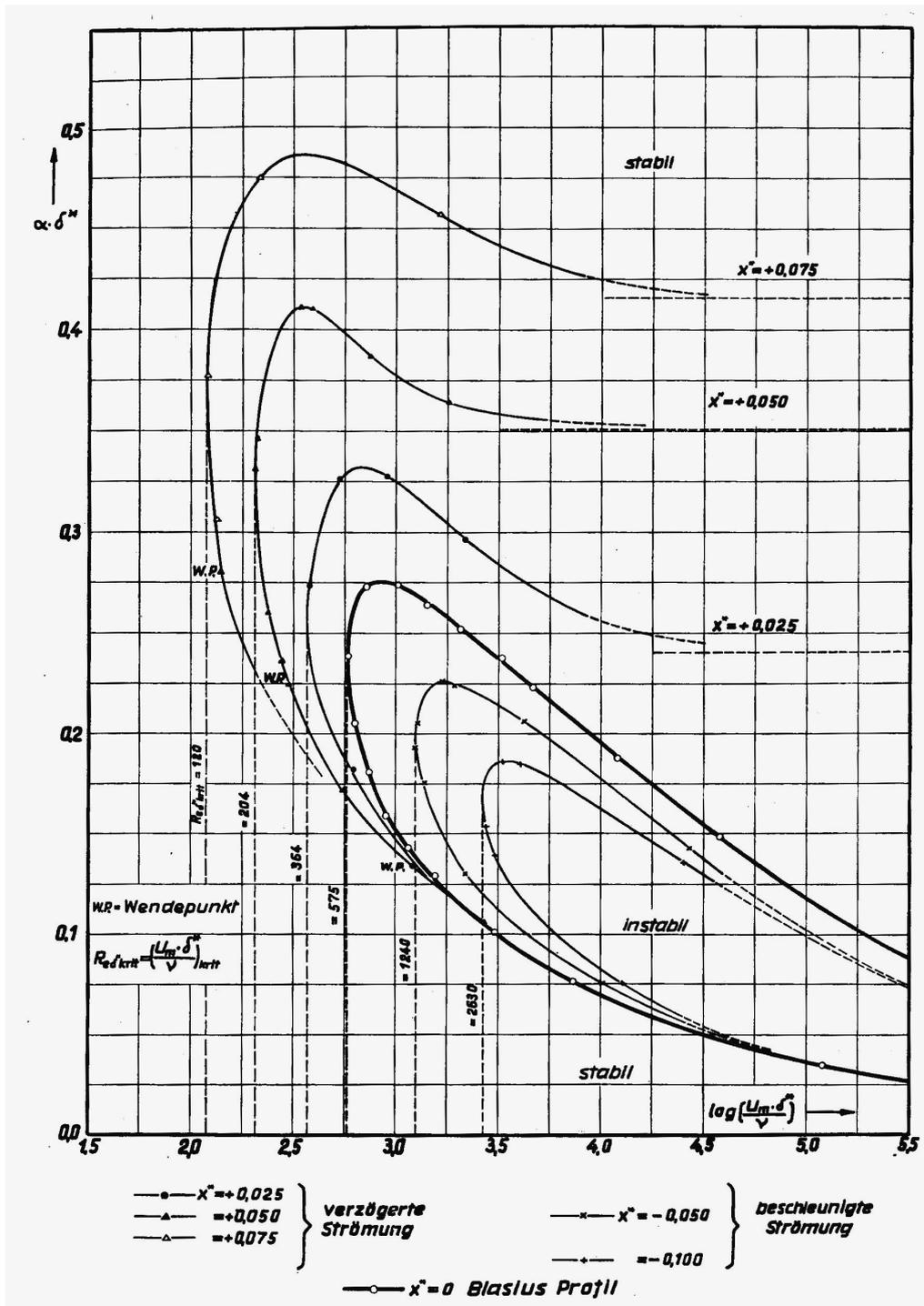

Fig. 3: This diagram shows curves of neutral stability, as derived from the Tollmien-Schlichting theory for boundary layer flow under various pressure gradients. The innermost curves correspond to flow conditions under favourable pressure gradients, such as with laminar airfoils, where the flow was accelerated by appropriately shaped profiles (Schlichting 1941: Fig. 11).

In December 1942, Schlichting's assistant presented in another war report the results of aerodynamic measurements of a number of laminar profiles performed in a wind tunnel of the LFA in Völkenrode. Certain theoretical profiles ("Joukowski-profiles") were compared with actually employed profiles with regard to their aerodynamic performance. The NACA profile 27215, for example, one of the favorite "laminar" profiles of the Langley aerodynamicists, was taken from a secret NACA report which they had obtained in French translation from the NACA's Paris office; it had been evacuated in a rush before the fall of France. Although it is not clear to what extent the measurements could reproduce the characteristics of these profiles, because the Völkenrode wind tunnel was not appropriate for low-turbulence testing, the virtues were clearly recognized. By comparison with a "normal" NACA profile (NACA 0010), the laminar profiles had a much lower profile drag (up to 50 percent) at small lift coefficients (Bußmann 1942). Other investigations were reported in January 1943 from the Deutsche Versuchsanstalt für Luftfahrt (DVL) in Berlin-Adlershof. Measurements in the large wind tunnel of this facility, too, confirmed that the Mustang wing had a considerably lower drag than other profiles of equal thickness. It was also found that with regard to the lift coefficient this profile had a "steady and relatively slow passage of the transition point" so that the drag minimum was "softly rounded and without jumps" (Doetsch 1943). Last, but not least, laminar profiles of Mustangs and other airplanes were subjected to wind-tunnel tests in various facilities. Schlichting's collaborator at the TH Braunschweig, Karl Bußmann, concluded in a war report in January 1943 from such measurements in two wind tunnels at the TH Braunschweig and at the LFA Völkenrode: "In the range of investigated Re-numbers up to 4,000,000, therefore, the profile P-51 'Mustang' behaves as a laminar profile." (Bußmann 1943: 5). Other tests made at the Aerodynamische Versuchsanstalt (AVA) in Göttingen provided more details about the behavior of such "laminar" profiles (Breford and Möller 1943; Riegels 1943). In 1943, a large low-turbulence tunnel at the AVA became operational, but it was apparently not used for laminar airfoil development (Holstein 1946). Although it is obvious that German aerodynamicists were fully aware of the potential of laminar wings, it is not clear what lessons the German aircraft industry drew from this knowledge.

**The Discovery of Tollmien-Schlichting Waves**

The pursuit of practical goals in aeronautics, such as laminar wings, involved the development of new techniques which ultimately led to the experimental verification of Tollmien's and Schlichting's predictions about boundary-layer instability. However, the corroboration for this theory came as a surprise. It did not result from a desire to confirm a scientific theory—a theory that in 1938 G. I. Taylor regarded as experimentally refuted—but from investigations performed in the course of war projects at the National Bureau of Standards (NBS). Throughout the 1930s and '40s, Dryden's Aerodynamics Division at the NBS pursued a number of projects under contract for the NACA. In May 1940, research was on the agenda about "Transition Phenomena at Low Turbulence", "Construction of Turbulence Measuring Equipment for Langley Field", and "Methods of Reducing Wind Tunnel Turbulence". The purpose of the first item on this list was to determine "the effect of small intensities of wind-tunnel turbulence on the position of transition from laminar to turbulent boundary-layer flow". A "Brief Description of Method" outlined how a flat aluminium plate was to be exposed to the air stream of the low-turbulence tunnel at the NBS; and the transition was to be studied "by surface pitot [i. e., direct speed measurement] and hot wire". Because other measurements had indicated that even at very low levels of wind-tunnel turbulence a transition could be evoked, "emphasis will be laid on reducing the turbulence as much as possible and investigating transition at the lowest values obtainable".[6]

Reduction of wind-tunnel turbulence, therefore, was the primary motivation, because otherwise it was not possible to arrive at a precise determination of the position of the laminar–turbulent transition in the boundary layer of a wing. Neither the NACA nor the NBS aimed at an experimental test of the stability theory. Galen B. Schubauer and Harold K. Skramstad, the experimenters in Dryden's group who performed these investigations, were experts in turbulence measurements by hot-wire anemometry with all its involved electronics. During hot-wire recordings from the boundary layer of the plate, they noticed that the oscillograph occasionally registered regular sinusoidal traces (Fig. 4).

---
[6] Lewis to Warner, 6 May 1940. Hugh L. Dryden papers, ms. 147, Special Collections, Milton S. Eisenhower Library, The Johns Hopkins University (Dryden papers), subject files, misc. correspondence, box 62.

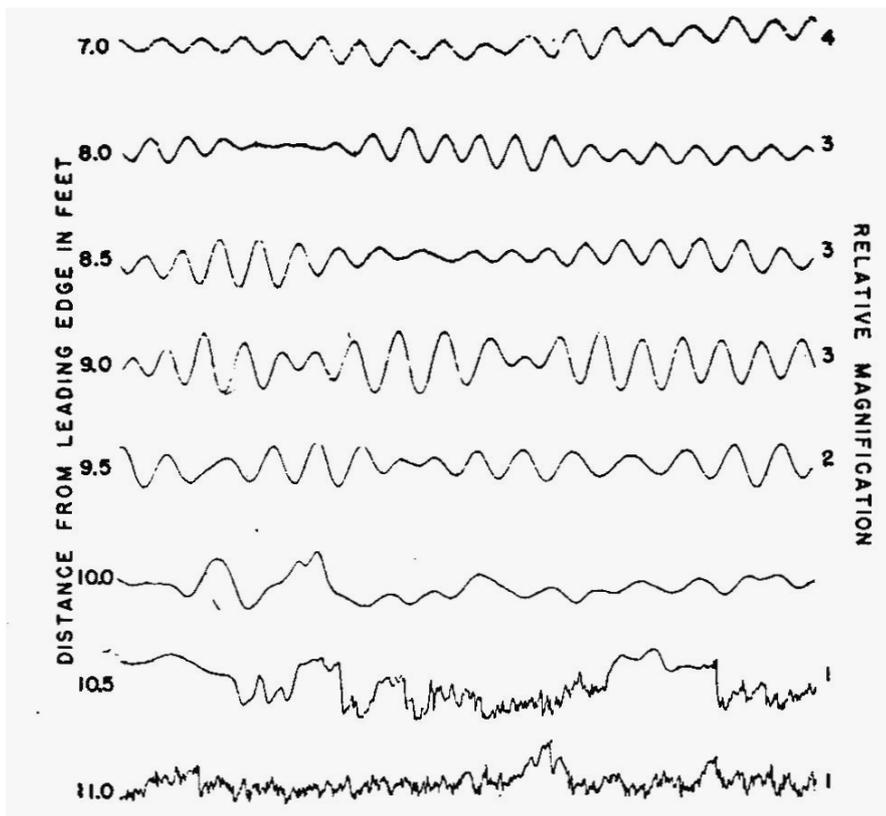

Fig. 4: Hot-wire oscillograph recordings of boundary layer oscillations at various distances from the leading edge of a flat plate; the distance of the hot-wire probe from the surface was 0.023 inch, the airspeed 53 feet per second (Schubauer & Skramstad 1947: Fig. 5)

The observation was reported in May 1941 as a rather surprising side-effect: "Under conditions of low stream turbulence, transition is found to result from growth in amplitude of a fairly regular oscillation of the boundary layer. The oscillation was in fact so regular that considerable work was done to eliminate the possibility that it might be due to mechanical vibration. The frequency was dependent on the speed of the stream and the thickness of the boundary layer. The oscillations have some of the characteristics of those predicted by the Tollmien-Schlichting theory which have not heretofore been experimentally observed. The frequencies are in fact in line with those predicted. Likewise the speed fluctuation in the outer portion of the boundary layer is in opposite phase to that near the wall as predicted by the theory." What later would be described as "Tollmien-Schlichting waves" was designated in this report as "a kind of boundary layer 'flutter'," alluding to the problems of wing "flutter."[7]

For the next fiscal year the common NBS-NACA research on transition phenomena at low turbulence was titled "Investigation of Laminar Boundary-Layer Oscillation." Only now the "natural oscillations of the boundary layer which have been studied by Tollmien and Schlichting" received full attention. But this does not mean that the research was now regarded from a more fundamental and less practical perspective: The investigation should take into account "pressure gradients of practical interest", the 1942 proposal argued, alluding to applications like low-drag wings where the boundary layer would not be in a zero pressure gradient.[8] The NACA pursued fundamental boundary-layer research at the height of the war only on condition that the results would be "applicable to current problems relative to military aircraft both to wing development and to ducting problems as well", headquarters advised the Langley Laboratory in February 1943 (Roland 1985, vol. 2: 549).

The laminar–turbulent transition in boundary layers was also a long-standing topic on the research agenda at the GALCIT in Pasadena, where Theodore von Kármán directed an ambitious group of theorists and experimenters. Here, too, these investigations were sponsored by the NACA because of the potential for aeronautical applications. In 1938, a small wind tunnel was built and adjusted to low-turbulence studies in a similar manner as Dryden's NBS tunnel. In 1942, Kármán entrusted a student, Hans Liepmann, with measurements on the stability of the laminar boundary layer on curved surfaces in this low-turbulence tunnel. Liepmann also found the oscillations predicted by the Tollmien-Schlichting theory. Both the NBS and GALCIT results were summarized in 1943 in Advance Confidential Reports for the NACA (Schubauer and Skramstad 1943; Liepmann 1943).

---

[7] Report to NACA, 10 May 1941. Ibid.
[8] Proposal to NACA for the Fiscal Year 1942. Ibid.

Thus, at the height of World War II, the disputed Tollmien-Schlichting theory was twice corroborated in the course of wartime research in American laboratories. A Chinese student of Kármán, Chia-Chiao Lin, added further evidence in the form of a theoretical investigation of the stability approach which became the subject of his PhD dissertation. Lin's revision, by and large, confirmed the disputed stability theories by Heisenberg on Poiseuille flow and Tollmien and Schlichting's on boundary-layer flow. Lin was at that time not yet a naturalized US citizen and thus not cleared for confidential war research. Therefore, unlike Liepmann's experimental investigation, his revision of the stability theory was not officially performed under a NACA war research contract. However, Lin thanked in the ackowledgment, among others, "Dr. H. W. Liepmann and several of his friends for many valuable discussions and suggestions." The knowledge of Liepmann's experiments would certainly have strengthened Lin's trust in his own findings; he noted that they "differ markedly from customary beliefs" (Lin 1944: 3). Whatever it was that Lin was allowed to know about the war research at the GALCIT, his doctoral advisor was now in a position to add the theoretical evidence to the recent experimental observations. Kármán sent Dryden a diagram "showing Lin's curve (present calculation) for the stability limit in the boundary layer compared to Schlichting and the experimental evidence." Lin's curve supported the conviction that Liepmann as well as Dryden's group had indeed observed the disputed Tollmien-Schlichting waves. "Unfortunately because of the confidential character of yours and Liepmann's work this diagram cannot be published", Kármán regretted. "However I thought you might be interested in having a copy for your file."[9]

**Postwar Reconstructions**

In 1946 Dryden disclosed these "Recent Contributions to the Study of Transition and Turbulent Boundary Layers" at the Sixth International Congress of Applied Mechanics in

---

[9] Kármán to Dryden, 26 February 1944. Dryden papers, folder "1935–1959. Misc. Dryden–Von Kármán Correspondence".

Paris. Subsequently it was oficially published in the *Journal of the Aeronautical Sciences*. Dryden labeled the phenomenon on this occasion "Tollmien-Schlichting Waves". The disclosure obviously came as a surprise. "Since the occurrence of the oscillations was unexpected, a number of tests were made to make certain that the oscillations were not merely some effect of vibration. The possibility of such effects was soon ruled out", the experimentators reported. The suspicion had been that the oscillations were an artefact of the measuring equipment, such as a vibration of the hot-wire "bug" with which the surface of the plate was scanned. Other possibilities, such as acoustic noise, were also considered as a source of the observed oscillations. But it soon became clear that the oscillations resulted from a selective amplification within a narrow band of frequencies. For further evidence that these oscillations were indeed those predicted by the Tollmien-Schlichting theory, they were excited artificially by a vibrating ribbon. By this method the frequencies amplified or damped at a given wind speed (i. e., Reynolds number) could be precisely determined. Finally, it became clear that the observed oscillations could only result from periodic velocity fluctuations caused by a wave that traveled downstream through the boundary layer (Dryden 1946, Schubauer and Skramstad 1947).

Germans were not invited to attend the Paris congress in 1946; but Tollmien and Schlichting probably soon learned about the experimental confirmation of their theories from British colleages at Farnborough, where they were working after the war under contract for the Royal Aircraft Establishment. In summer 1947, Tollmien was called to Göttingen as Prandtl's successor on the chair of applied mechanics at Göttingen University. Dryden congratulated Tollmien "on your appointment to succeed Prandtl at Göttingen" and used the opportunity to send him a comprehensive account about the paper by Schubauer and Skramstad: "You must feel proud that your early theoretical work has been so fully confirmed. The real factor making the work possible was the attainment of an air stream of low turbulence."[10]

Tollmien, of course, was pleased to hear this news. "I had never hoped to see such a satisfactory confirmation of the stability theory", he responded.[11] However, he was far from enthusiastic about Lin's theoretical paper which was published in 1945 and 1946 in the subsequent issues of the *Quarterly of Applied Mathematics* (Lin 1945 and 1946). When Lin heard rumors about Tollmien's objections and asked for the reasons, Tollmien detailed in a five-page letter what upset him. In particular, he criticized Lin's version of the historical development: "The history of the stability research of laminar flows is rather complicated", he began his critique. He was most upset that Lin created the impression that Tollmien had merely adopted Heisenberg's approach. The most important first step had been taken by Prandtl and Tietjens, before Heisenberg, Tollmien explained, so the priority for opening the route towards calculations of stability curves belongs to Prandtl's school. "Heisenberg was, during the early twenties, very often at Göttingen and was completely familiar with the ideas current at this time in the Prandtl school on the stability problem." Tollmien further criticized that Lin had not mentioned Noether's "frightening criticism" which dealt an "almost deadly blow" to the whole approach. He also distanced his own work from Heisenberg's "doubtful expansions" and claimed his own priority for having elaborated the crucial physical concept of the "inner friction layer". This resulted in the first calculation of a curve of neutral stability and served as a starting point for other related work in Prandtl's school. "Summarizing the whole situation, I think it is really silly to suppress my name in the history of the stability problem in favor of Heisenberg's or yours as it has been done in several American publications."[12]

---

[10] Dryden to Tollmien, 27 August 1947. Archive for the History of the Max-Planck-Gesellschaft, Berlin (MPGA), div. III, rep. 76B, Walter Tollmien papers, box 1 (correspondence 1947–1950).
[11] Tollmien to Dryden, 9 April 1948. Ibid.
[12] Tollmien to Lin, 24 November 1948. Ibid.

In his response, Lin did not defend his flawed historical reconstruction, but he played down the difference between the methods, by which Tollmien characterized Heisenberg's work as secondary to that of Prandtl's school. Lin insisted that Heisenberg's series expansion in the inviscid case was "valid and led to the same results as yours, while you registered some doubt on those series". Nevertheless, Lin finally gave in with the remark: "Since you and others think that Heisenberg's expansion method and your method are different, I must admit that you did not 'adopt' a well established method. I must apologize to you for using that word."[13]

But the controversy did not end with Lin's lukewarm apology. In April 1949, Tollmien forwarded a paper to the editor of the *Zeitschrift für Angewandte Mathematik und Mechanik* (*ZAMM*) written by Horst Holstein, the AVA's expert on low-turbulence research during the war (Holstein 1950). The article opposed Lin's historical reconstruction, and Tollmien urgently wanted his criticism to become widely known. As he explained: "In 1945 many foreigners believed that German science was dead and that one can get away with plundering in our garden".[14]

Lin felt "really puzzled" about this turn of events. "On the historical point, I still feel that I should reserve my final comments", he wrote to Tollmien in June 1950. Regarding the critique of other parts of his work, he believed that "Holstein completely missed the point".[15] Tollmien responded with praise for Holstein as a "physicist with great experience in theoretical and experimental aerodynamics. He is working now in England. When he wrote his paper he was a member of our institute." After repeating the main points of dissent about the origins of stability theory, he added in a conciliatory tone: "In spite of the fact that you have done grave injustice to me in your paper I do not bear any grudge against you because I think you were ill advised. You can recognize my mentality from the patience with which I explained the historical facts to you for the second or third time." But he did not miss pointing out that he might have other weapons at his disposal: "I have refrained from using the opportunity to oppose you in an American periodical which has been offered to me by an American aerodynamicist of highest rank."[16]

---

[13] Lin to Tollmien, 7 June 1949. Ibid.
[14] Tollmien to Willers, 4 April 1949. Ibid.
[15] Lin to Tollmien, 30 June 1950. Ibid.

"The fateful Mr. Lin has, of course, responded and addressed a quite outraged letter to me", Tollmien subsequently informed Holstein. Schlichting was "even more upset than I about Lin's behavior", he added.[17] Tollmien also informed Heisenberg about the controversy. In view of Lin's praise of Heisenberg's dissertation as the major breakthrough for the asymptotic stability theory, it is not surprising that Heisenberg had a more favorable opinion of Lin's work. However, as Tollmien wrote in his response to Lin, after "a long talk" with "my dear colleague Heisenberg, who has his institute next door to mine" there was "no disagreement on this matter".[18] But from his correspondence with Heisenberg it becomes obvious that total agreement on this matter it was not. Tollmien found it necessary to explain in a rather defensive tone why he mentioned Lin's stability theory in his forthcoming FIAT review (Tollmien 1953) only "by way of a parenthetical"; he admitted that Lin is "a promising young scientist, who was only somewhat poorly advised in 1945 to tackle a task whose difficulty he did not oversee".[19]

In other words: Tollmien discerned Kármán, Lin's thesis advisor, as the true culprit. Kármán learned about the controversy through his correspondence with Lin. "I did not see the paper of Holstein", Kármán wrote to Lin, "in any case I think I should attack him, in your place, if he attacked you."[20] But the dispute did not escalate further, as new discoveries about the onset of turbulence in boundary layers funneled the interest into new areas of research.

---

[16] Tollmien to Lin, 24 July 1950. Ibid.
[17] Tollmien to Holstein, 7 August 1950. Ibid.
[18] Tollmien to Lin, 24 July 1950. Ibid.
[19] Tollmien to Heisenberg, 18 July 1950. Ibid.
[20] Kármán to Lin, 5 July 1950. Theodore von Kármán Collection (TKC), Archives of the California Institute of Technology, Pasadena, 18-25. (Microfiche copies are available also at the Smithsonian Air and Space Museum in Washington, D.C., and the Institut für Philosophie und Wissenschaftsgeschichte der Universität Bern.)

## "Turbs" and "Horseshoes"

By the early 1950s the study of the onset of turbulence was causing new excitement about experimental research rather than theoretical considerations. It was clear to boundary-layer experts even before the war that the limit of stability was not identical with the onset of turbulence. The initial instability could trigger a transition, but the final onset of turbulence involved processes by which the plane Tollmien-Schlichting waves had to break up before the motion would become turbulent. "One imagined that the transition finally happens abruptly along an irregular up-and-down-moving line", Henry Görtler, another Prandtl disciple, described the current opinion in the early 1950s (Görtler 1955). The discovery of so-called "Emmons's spots" or "turbulent spots" provided a first glimpse into the complexity of this process (Emmons 1951). Howard Wilson Emmons, a professor at the Harvard Engineering School, had been involved during the war in research about supersonic flow and shock waves. In 1945 he had participated in a Naval Technical Mission, which evaluated German war research in these areas. "Supersonic flow was in the forefront of new aircraft and missile problems", Emmons later recalled. In order to visualize supersonic phenomena he used a simple water channel. The wave crests around an obstacle in the shallow flow of "shooting" water (i. e., when the water flows faster than the speed with which water waves propagate on its surface) display similar characteristics as supersonic flow phenomena. Emmons intended to use this so-called "hydraulic analogy" as a laboratory demonstration in his institute at Harvard University. "The demonstration worked just great", he recalled. Occasionally, however, he observed "peculiar wiggles" which passed downstream. Emmons suspected that this phenomenon played a role in the transition to turbulence and called the wiggles "turbs". He offered the phenomenon as a topic for a doctoral thesis to a physics student, Morton Mitchner. When Hugh Dryden learned about Emmons's "turbs", he "dismissed the whole thing as merely a surface wave phenomenon of some sort having no connection with transition". Perhaps in an attempt to refute Emmons, Dryden asked Schubauer and another member of the Aerodynamics Section of the NBS, Philip S. Klebanoff, to look for this phenomenon in the boundary layer on a flat plate in the NBS low-turbulence tunnel. "I'm sure that Schubauer and Klebanoff were told that they probably would find no turbulent spots because there was no free surface", Emmons mused in retrospect.[21]

Dryden had been director of the NACA since 1947. He was no longer involved actively in fluid dynamics research at the NBS. However, as testified by his frequent review articles, he was keenly interested in the investigations on boundary-layer transition in his former laboratory, where Schubauer was pursuing the research tradition that Dryden had started thirty years ago. Klebanoff, who had joined the NBS aerodynamics section in 1942 as a young physicist, inherited the major responsibility for experimental transition studies in the low-turbulence tunnel when Schubauer took over the leadership of the laboratory in 1947. Emmons's observation of "turbs" naturally attracted Schubauer and Klebanoff's attention, whatever Dryden may have thought about their nature. In February 1955, they presented a comprehensive report of their findings. They used the same experimental arrangement as Schubauer and Skramstad for their study of Tollmien-Schlichting waves in the early 1940s: A 12-foot-long aluminium plate with a sharpened leading edge spanned across the 4-½-inch-wide test section of the NBS low-turbulence tunnel; they investigated both natural transitions and artificially induced transitions, for example, by electric sparks initiated by a needle electrode placed a ¼ inch above the surface (Fig. 5). As a result, they largely confirmed that Emmons's "turbs" were indeed a phenomenon related to the onset of turbulence and not a free surface effect as Dryden had first guessed (Schubauer and Klebanoff 1955).

---

[21] Emmons, "Some Fluid Mechanic History." Manuscript of a speech on the occasion of the Award of the Fluid Dynamics Division of the American Physical Society, 29 March 1982. Archives of the Division of Fluid Dynamics of the American Physical Society, Lehigh University, Bethlehem, (DFD-Archives), 13-02.

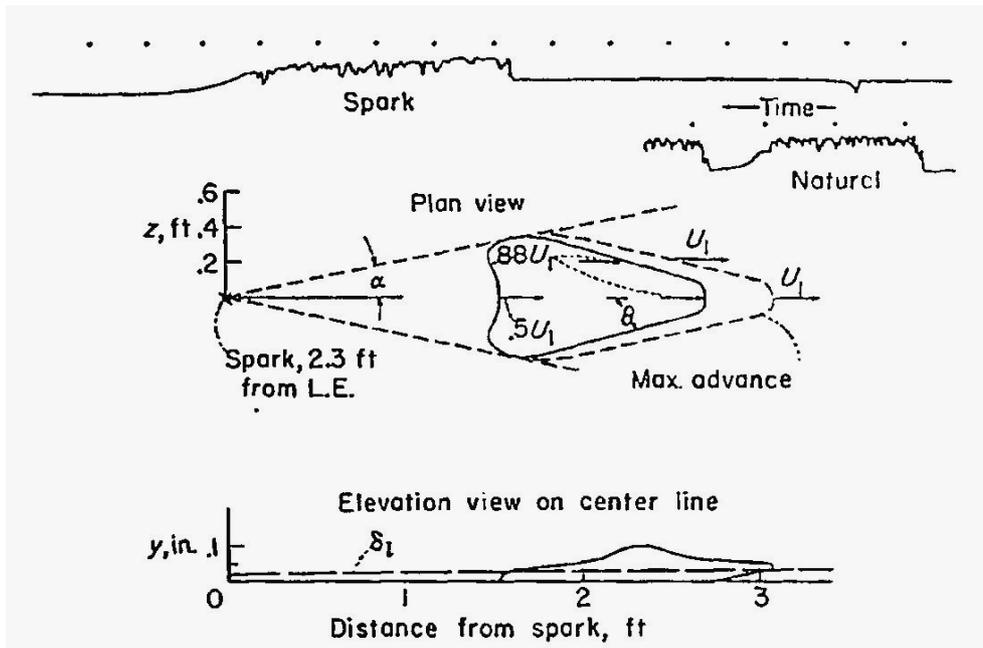

Fig. 5: Turbulent spot in a laminar boundary layer on a flat plate. The upper oscillogram traces the passage of spark induced and naturally occuring spots (the distance between two points corresponds to 1/60 second); underneath are shown plan and elevation views of spark-induced turbulent spots. The airspeed U1 was 30 feet per second (Schubauer & Klebanoff 1955: Fig. 6).

In August 1957, the International Union of Theoretical and Applied Mechanics (IUTAM) convened a symposium on "Boundary Layer Research" in Freiburg in Germany, where the recent investigations on stability and transition were reviewed. Schubauer reported about work in progress at the National Bureau of Standards. "The initial objective", he explained "was to study phenomena at the instantaneous point of change-over from laminar to turbulent flow". The experiments focused on the final stage of the transition before the boundary layer became fully turbulent as a result of turbulent spots being swept downstream as they grew and merged with one another until no trace of laminar flow was left in the transition region. It turned out that the emergence of spots was related to the amplification of Tollmien-Schlichting waves, but it was unclear how discrete patches could result from plane waves. Therefore, "the objective was to examine more completely the connection between the wave and transition". Thus was Schubauer's explanation for the shift in emphasis from the final to the intermediate stage of the transition process. He presented a diagram showing how the amplitude of the Tollmien-Schlichting waves was modulated laterally farther downstream: the wave intensity measured by the hot-wire probe was no longer constant when it moved at a right angle to the flow direction over the plate. Gradually the hills and valleys steepened into peaks and troughs. The outburst of turbulent spots "never occurred without first being preceded by a wave strongly warped into peaks and valleys", Schubauer summarized the result of these measurements (Schubauer 1958). The warped wavefront seemed to degenerate into vortex loops, such as observed by other experimenters in water flow over a flat plate before the outbreak of turbulent spots. These loops reminded Schubauer of an idea that the former NACA theorist, Theodore Theodorsen, now at the Fairchild Engine and Airplane Corporation, had recently presented: According to Theodorsen, "Tollmien-Schubauer waves" gave birth to "horseshoes", a shape of vortices which he regarded as the universal elements of turbulence (Theodorsen 1956).

In 1962, Klebanoff and two other members of Schubauer's group published a comprehensive review of their experimental investigations of the transition from laminar to turbulent flow in the boundary layer along a flat plate. They now discerned three stages of the transition process, "a primary stage which is governed by the two-dimensional linearized stability theories, a second stage of finite amplitude where strong three-dimensional effects are observed, and a third stage involving the birth of turbulent spots". Their focus was on the latter two stages, because they considered the initial phase of the transition process as "fairly well" established by the earlier investigations of their group. Using the same vibrating-ribbon technique for exciting controlled Tollmien-Schlichting waves on the same flat aluminium plate in the same low-turbulence tunnel with which Schubauer and Skramstad had started boundary-layer transition investigations at the NBS twenty years ago, Klebanoff and his collaborators had more experience than any other research group for tackling the experimental problems. They surveyed the onset of turbulence within the minute boundary layer—particularly where it became important to register its three-dimensional structure. In order to trigger controlled spanwise modulations of the wavefront of Tollmien-Schlichting waves, they placed 0.003-inch-thick strips of cellophane tape distanced at 0.5 inches from one another on the plate underneath the vibrating ribbon. Their hot-wire measurement equipment was precise enough to register air speeds at heights over the plate that differed by hundredths of an inch, and this precision was enough to record the streamlines in vertical cross sections of the boundary layer. Thus the deviations from the laminar (Blasius) velocity profile downstream with increasing distance from the vibrating ribbon could be associated with the superposition of a longitudinal eddy system. Farther downstream, Klebanoff and his collaborators analysed what they described as "breakdown": The oscillogram recorded from the hot-wire probe at such positions no longer displayed a wave-like trace but spikes, which they regarded as a "manifestation of 'hairpin' shaped eddies", analogous to the structures observed by other experimenters in the unsteady flow behind pointlike obstacles (Fig. 6). In order to make this analogy more apparent they cemented a hemispherical obstacle with a radius of 0.063 inch on the plate two feet behind the leading edge (where it reached the critical depth of the boundary layer from where the breakdown originated). The instantaneous velocity distribution measured at the site where the spikes occurred showed the same profile as the mean velocity distribution in the wake downstream of the obstacle. Both had a significant inflection at the same critical depth, hinting at the same sort of inflection instability. Thus, the debate circled once more around instabilities (Klebanoff et al. 1962).

**The Rise of Nonlinear Theories**

Tollmien-Schlichting waves resulted from a two-dimensional theory: Spanwise variations were not taken into account. The instability having to account for these variations involved the third spatial dimension. Three-dimensional effects in linear stability theory had been neglected so far because "Squire's theorem" stated that two-dimensional disturbances are more unstable than three-dimensional disturbances (Squire 1933). Nevertheless, the amplification and warping of Tollmien-Schlichting waves, the emergence of longitudinal vortical motion, the "horseshoes" and "hairpins", and the appearance of turbulent spots were obviously three-dimensional effects. The streamwise vortices resembled "Görtler vortices", resulting from another instability which arose in flows along concave walls. It was extensively debated at the IUTAM symposium in Freiburg in 1957 whether a secondary instability of this type could explain the counter-rotating longitudinal vortices in the transition zone (Görtler and Witting 1958). Concave streamline curvature, according to this instability, would also result in vortex formation. But the longitudinal vortices did not originate from the valleys of the warped wavefronts, where the curvature of the streamlines was concave, but from the peaks, where it was convex. Klebanoff and his colleagues therefore concluded, "that the Görtler-Witting mechanism does not provide an adequate explanation for the observed behaviour" (Klebanoff et al. 1962: p. 16).

  Lin regarded the lack of an explanation as a challenge and an opportunity to claim a leading role in this area. In the 1950s, he ambitiously raised his own research school at MIT in the "GALCIT spirit", as he wrote to his former teacher in March 1957. He revealed that he was particularly focusing on "the problem of oscillations of finite amplitudes in a laminar flow. A group of us are now working on such problems."[22] One of this group was David J. Benney. In 1959, Benney finished his Ph.D. under Lin's supervision and subsequently collaborated with him on the stability theory. They were also in close contact with the experimental researchers in Schubauer's laboratory at the National Bureau of Standards. In April 1960 Lin reported about recent progress which he and Benney had made along the lines suggested by Kármán, namely to extend the stability theory by keeping the nonlinear terms in the equation of motion for the perturbed laminar flow. "We tried it. At first it appeared intractable, but we finally found a very elegant way to classify the nonlinear effects." In

---

[22] Lin to Kármán, 12 March 1957. TKC 18-27.

contrast to Görtler and Witting, Benney "obtained agreement with Schubauer's experimental results".[23]

A month later, in May 1960, Benney and Lin sent a preliminary report to *Physics of Fluids*. Their theory analyzed the interaction of a plane Tollmien-Schlichting wave with a superimposed three-dimensional wave with a spanwise varying amplitude. A perturbation of a plane flow with such a combination of waves gave rise to a longitudinal vortical motion which was "strongest at points where the primary oscillation yields a convex streamline, in contrast to the predictions of the Görtler-Witting theory". Benney and Lin used a mathematically more tractable velocity profile of a plane shear flow, $U(y) = \tanh(y)$, which was symmetric to a central plane at $y = 0$, rather than the Blasius profile of laminar boundary-layer flow along a flat plate. This choice made the comparison with the boundary-layer experiments at the NBS problematic because the inflection at $y = 0$ had already resulted, with the linear approximation, in a different stability behavior. Nevertheless, or rather because of this difference, they found the agreement between theory and experiment "all the more remarkable" and argued "that this indicates the generality of the features revealed by the present investigations" (Benney and Lin 1960). A few years later, Benney also treated a basic flow profile that was closer to the Blasius profile for laminar boundary-layer flow—and arrived at a similar conclusion. Both in free-shear flow and in boundary-layer flow the nonlinear interaction of a plane (Tollmien-Schlichting) wave with a wave moving in a spanwise direction results in a longitudinal vortical motion (Benney 1964, Betchov and Criminale 1967).

The nonlinear interactions of different modes of perturbation were not the only source of instability beyond the reach of linear theory. Nonlinear effects had already arisen with a single disturbing wave when the amplitude was finite (and no longer assumed to grow from infinitesimally small values). Such a nonlinear effect was supposed to be responsible for the empirical observation that finite disturbances could make the laminar flow unstable far below the Reynolds number for which the linear approach predicted instability. In 1951, two mathematicians from the Imperial College of Science and Technology in London, David Meksyn and John Trevor Stuart, published a paper on the stability of plane Poiseuille flow for finite disturbances. If infinitesimal disturbances were assumed as the only source of instability, the lowest critical Reynolds number would not display a dependency on the amplitude of a disturbing wave. But Meksyn and Stuart, by taking nonlinear terms into account, found that the lowest critical Reynolds number was indeed dependent on the magnitude of the disturbance. According to their calculation, the critical Reynolds number drops with increasing amplitude of the disturbing wave; it reaches a minimum of 2,900 at some finite amplitude—in contrast to a value of 5,050 for infinitesimal disturbances as predicted by the linear theory (Meksyn and Stuart 1951). During the subsequent years, Stuart further elaborated the nonlinear theory. He used "Burgers's model" (Burgers 1948), a nonlinear one-dimensional equation introduced originally as a simplified model for the Navier-Stokes equations, in order to illustrate the possibility of a "limit-cycle" solution (Stuart 1956). In 1957, he also introduced a new vocabulary in order to characterize the peculiar differences between the linear and nonlinear theories. He called a flow "subcritical" if it was stable for infinitesimal disturbances and unstable for finite disturbances. If a finite disturbance would not permanently destabilize the flow but lead to a new equilibrium, Stuart called it "supercritical". In the course of his analysis, Stuart rediscovered an equation for the square of the amplitude of a disturbance that had been studied first by Lev Landau in 1944. The "Landau equation" became known internationally in 1959 through the translation of the famous Landau-Lifschitz textbook on fluid dynamics (Landau and Lifschitz 1959). Landau had postulated this equation as a general description of flow behavior in the neighborhood of a critical Reynolds number; Stuart's "Landau equation" addressed a special case, and thus confirmed Landau's vision of how to approach fluid mechanical instabilities in terms of nonlinear theories (Stuart 1958).

---

[23] Lin to Kármán, 13 April 1960. Ibid.

Based on these early "weakly nonlinear" approaches, as they were later labeled, the theory of hydrodynamic stability developed into a multitude of concepts and methods. Just as Klebanoff and his coworkers' hot-wire oscillograms from the low-turbulence tunnel at the National Bureau of Standards exposed the complexity of the transition region between laminar and turbulent flow from an experimental vantage point, theoreticians became aware how nonlinear approaches enriched the understanding of the onset of turbulence. Stuart recalled this period many years later as "the expansive 1960s" (Stuart 1986). Nonlinear theories were flourishing. However, as the author of another review article observed with hindsight in 1988, research on boundary-layer transition was always prone to "sudden changes in perspectives and trends". Even review articles on such a narrow research field as shear-flow instability had "little in common and a zero-overlap of cited literature". Only the very first stage of the transition process, when Tollmien-Schlichting waves emerged as a consequence of a primary instability, appeared well enough understood for widespread consent among expert opinion. However, compared to the complexities of the subsequent stages, they now seemed almost trivial, if not irrelevant, because they could easily be bypassed by external causes of the type G. I. Taylor had in mind when he had opposed the Tollmien-Schlichting concept in the 1930s: "Because of the multitude of identified and concealed effects on the response of the boundary layer to external forcing, the development of a general theory of transition is yet a utopia" (Herbert 1988; for other reviews and opinions see Craik 1985, Kachanov 1994, Yaglom/Frisch 2012).

**Conclusion**

The frustrated theories on turbulence call for answers about the underlying causes. But even in the limited case examined by this study, the onset of turbulence in a flow along a flat plate, it is problematic to discern one or another "external" or "internal" cause to explain delays or breakthroughs. Instead of answers we are left with more questions calling for more detailed case studies. Within the scope of this study, specific answers were obtained by closer inspection of Prandtl's, Dryden's, and von Kármán's institutes, such as the impact of military causes in World War II (laminar airfoils, wind-tunnel turbulence), but this is not sufficient to warrant any conclusion about the military impact in general.

Nevertheless, this study brought to the fore some features that deserve broader attention. Most striking, from an epistemic perspective, is the role of the wind tunnel—a device known to serve primarily engineering goals. While the wind tunnel figures strongly in the history of aeronautical technology, it is mentioned only cursorily, if at all, in reviews of fluid-mechanical phenomena. These are traditionally labeled "scientific" rather than "technological", such as turbulence as one of the last unsolved "basic" riddles of classical physics. Although this study was not pursued with an emphasis on theory–experiment relations in fluid mechanics, it should have become clear that the investigation of low-turbulent flow in special wind tunnels during World War II was crucial—not only for the discovery of the Tollmien-Schlichting instability in particular, but also for turbulence research in general. G. I. Taylor's argument against the stability approach as pursued by Prandtl's school was nourished by wind-tunnel measurements undertaken within the context of his statistical turbulence theory in the mid-1930s. The same theory, motivated by the problem of wind-tunnel turbulence, marked the beginning of more "basic" theories on fully developed turbulence in the 1940s, such as those by Andrei Nikolaevich Kolmogorov and others. This strand of the "turbulence problem" bundle, therefore, also had major roots in the wind tunnel (Eckert 2008b).

The "bi-perspective" character of the wind tunnel as an engineering tool and a research instrument, usually installed and operated within the same institutional setting, has important consequences also for the theories and research programs involved. Due to this ambiguity, fluid-mechanical problems such as the "Tollmien-Schlichting instability", or more generally "boundary-layer theory", may be portrayed as "scientific" by exposing their pertinence to fundamental studies on turbulence, or as "applied" wind-tunnel research—and presented in one or the other orientation to different civil or military agencies for fundraising purposes. Entire institutions may adapt their research according to external priorities by exposing one or the other perspective. Such reorientations have been observed, for example, in Prandtl's Kaiser-Wilhelm-Institut für Strömungsforschung during the "Third Reich" (Epple 2002). By the same token, military agencies like NATO's AGARD fostered "fundamental" fluid mechanics and "applied" aeronautics. To those involved, like Dryden in his capacity as director of the NACA, this has become an almost trivial insight, "for what seems basic or fundamental to one group is regarded as applied by another" (Dryden 1954: 118).

The ambivalence between "applied" and "basic" is not limited to the role of the wind tunnel but a recurrent feature of twentieth-century fluid dynamics (Eckert 2006). When and to what extent one or the other tendency prevails is contingent on the social context. The NACA, for example, found it expedient in 1951 to translate Heisenberg's doctoral dissertation and, despite its apparently fundamental character, make it available to the clientele of its *Technical Memorandum* series of reports (Heisenberg 1951). The same contingency makes it difficult to estimate the mutual impact of cognitive and social aspects upon the development of turbulence research. What appears striking from one vantage point becomes obvious from another: Lin's corroboration of the discredited Orr-Sommerfeld approach during World War II would seem accidental from its "basic" conceptual premises only, because no new mathematical concept or physical insight was used that would not have already been available earlier; the historical reconstruction becomes plausible, however, when the "applied" context of laminar airfoils, wind-tunnel turbulence and the ensuing wartime research of the NACA with its programs at Dryden's and von Kármán's institutes is taken into account.

As a major lesson from this study I therefore conclude that such labels as "pure" and "applied" are intrinsically problematic and their use can only be made intelligible by reconstructing the historical context of the research to which they are attached. For example, an analysis of the dispute between Tollmien and Lin over theories of hydrodynamic stability, demands an awareness of diverse institutional legacies, embodying different local research traditions, loyalties, and perspectives. The mix of "applied" and "pure" research efforts cannot be disentangled, even retrospectively to assign one or the other label as the predominant attribute. Sommerfeld's Munich school of theoretical physics, Prandtl's Göttingen school, Kármán's group at the GALCIT, Lin's rising school at the MIT—each of these represented specific research traditions and environments, sometimes in close adherence to one another (such as Lin's attempt to implant the spirit of the GALCIT at MIT), but more often in-between, a mixture between adherence and rivalry (such as Kármán's GALCIT, in its relation to the Prandtl legacy).

The self-allotted attribute "applied" in the designation of journals and organizations (ZAMM, GAMM, IUTAM), which provided an institutional umbrella for the fledgling specialty of hydrodynamic stability research since the 1920s, hints at a certain self-image which many researchers in the field seem to have shared. Nevertheless, hydrodynamic stability and turbulence have never lost their character as "fundamental" mysteries—despite the practical interests involved. The great "Kelvin, Rayleigh, and Reynolds" struggled in vain to explain the onset of turbulence, remarked one reviewer in 1993, and "many of the presumably simpler phenomena of hydrodynamic stability also remain incompletely understood" (Trefethen et al. 1993: 578). When discussed within the context of the "turbulence problem", hydrodynamic stability research gains the aura of one of the ultimate and deepest riddles of the universe. The same observation can be made with reviews concerning theories of fully developed turbulence (Eckert 2008a).

The "applied" self-image of fluid mechanics and the mystification of coping with "fundamental" last riddles appear contradictory at first sight. But it is inappropriate to the concerned scientific community of fluid dynamicists as much as to its subject matter to analyse its disciplinary features along the applied-versus-basic divide. Fluid dynamics is a discipline without a clear-cut disciplinary structure. Investigations of fluid mechanical instabilities have been—and still are—performed within the traditional scientific disciplines of physics, mathematics, and various engineering specialties; however, in the course of their historical development, disciplinary boundaries have been transgressed as opportunities or needs arose. In view of this complexity it is problematic to generalize the conclusions from one case to another. Scientists with an interest in the Tollmien-Schlichting instability usually had a close connection with aeronautical engineering. Other types of instability, like the Rayleigh-Bénard instability (caused by thermal differences) or Taylor-Couette instability (caused by centrifugal forces), imply ramifications in other disciplinary contexts (geophysics, meteorology, astrophysics). It is this diversity of contexts which renders the history of fluid dynamics so complex, even when restricted to narrower specialties like flow instabilities and turbulence.

**Acknowledgments**

I thank the organizers of the Rauischholzhausen workshop for creating the productive atmosphere of exchanging work-in-progress on the history of fluid dynamics, and for the ensuing critical comments which accompanied the completion of this article. I am particularly grateful to David Bloor for polishing this article and many stimulating discussions.